\documentclass[aps,pra,twocolumn,amsmath,amssymb]{revtex4} 
\usepackage{epsfig}
\begin{document}  
\title{Violation of self-similarity in the expansion of a 1D Bose gas}

\author{P. Pedri$^{1,2}$, L. Santos$^{1}$, P. \"Ohberg$^{3}$ and S. Stringari$^{2}$}  
\address{(1) Institut f\"ur Theoretische Physik, Universit\"at Hannover, D-30167 Hannover,Germany}
\address{(2) Dipartimento di Fisica, Universit\`a di Trento and BEC-INFM, I-38050 Povo, Italy}
\address{(3) Department of Physics, University of Strathclyde, Glasgow G4 0NG, Scotland}

\begin{abstract}  
The expansion of a 1D Bose gas is investigated employing the Lieb-Liniger equation of state within the 
local density approximation. We show that during 
the expansion the density profile of the gas does not follow a self-similar solution, as one would 
expect from a simple scaling Ansatz. We carry out a variational 
calculation, which recovers the numerical results for the expansion, 
the equilibrium properties of the density profile, and the frequency of the lowest compressional mode. 
The variational approach allows for the analysis of the expansion in all 
interaction regimes between the mean field and the Tonks-Girardeau limits, 
and in particular shows the range of parameters for which the expansion 
violates self-similarity.
\end{abstract}  
\maketitle


\section{Introduction}


The experimental achievement of Bose-Einstein condensation (BEC) \cite{BEC} has aroused 
a large interest in the physics of ultracold atomic gases. 
Among the topics related to this field, the physics of low dimensional atomic gases has 
recently attracted significant attention. 
The development of the trapping techniques has allowed for the realization of 
very anisotropic geometries, 
where the confinement is so strong in one or two dimensions, that at 
low temperatures the transversal motion is ``frozen'', 
and does not contribute to the dynamics of the system.
In this way two- \cite{MIT,Safonov,Hansel,Florence} and one-dimensional \cite{MIT,Salomon,Greiner1} 
systems have been accomplished. Low-dimensional gases 
present significantly different properties compared to the three-dimensional ones. A remarkable example 
is provided by the existence of quasi-condensation \cite{Petrov1D,Kagan,Petrov2D,Petrov3D}, whose effects 
have been recently observed experimentally \cite{quasiBECexp}.


During the last years, the 1D Bose gases have been the subject of growing interest, in particular 
the limit of impenetrable bosons \cite{Girardeau60}, which behave to a large 
extent as a noninteracting Fermi system, acquiring some remarkable properties. 
The conditions for the experimental 
realization of strongly correlated 1D gases 
are rather restrictive \cite{Petrov1D,Olshanii1}, since a large radial compression, 
a sufficiently small density, and eventually a large scattering length are needed. Fortunately, recent 
experimental developments have opened perspectives in this sense. Especially 
interesting is the possibility to modify at will the interatomic interactions by means of 
Feshbach resonances \cite{Cornish}, and the capability of loading an atomic gas in an optical lattice \cite{Greiner}.


From the theoretical side, the physics of 1D Bose gases was first investigated by 
Girardeau \cite{Girardeau60}, who  
considered the limit of impenetrable bosons, also called Tonks-Girardeau (TG) gas,  
pointing out a non trivial relation with the physics of ideal Fermi gases. 
This analysis was later extended by Lieb and Liniger \cite{Lieb63}, who solved 
analytically the problem for any regime of interactions, using Bethe Ansatz. 
Yang and Yang \cite{yang} extended the analysis including  
finite temperature effects. Recently, the experimental accessibility of trapped gases, have encouraged the 
investigation of the harmonically trapped case. 
The Bose-Fermi (BF) mapping has been employed to the case of 
an inhomogeneous gas in the TG limit \cite{Wright2}. However, 
there is unfortunately, to the best of our knowledge, no exact solution for arbitrary  
interaction strength in the case of trapped gases. 
The problem of the equilibrium of a trapped gas can be analyzed using a local density
approximation and employing the Lieb-Liniger (LL) equation of state locally  to evaluate the equilibrium
density profiles \cite{Olshanii2}. A similar formalism has been recently employed to analyze the 
collective oscillations in the presence of harmonic trapping \cite{Chiara}. 
Both Refs.\ \cite{Olshanii2} and \cite{Chiara} have shown the 
occurrence of a continuous transition from the mean field (MF) regime to the TG one as the
intensity of the interaction is varied. 
Recently, Gangardt and Shlyapnikov \cite{Gangardt} have discussed the 
stability and phase coherence of 1D trapped Bose gases. These authors have 
analyzed the local correlation properties and found that inelastic decay processes, 
such as  three body recombination, are suppressed in the TG regime, and intermediate regimes between MF and TG.
This fact opens promising perspectives towards the accomplishment of strongly interacting 1D Bose gases 
with large number of particles. This analysis have been very recently extended to the case of finite temperatures 
\cite{Kheruntsyan}.

The expansion of a one-dimensional Bose gas in a guide was analyzed in Ref.\ \cite{Letter}, 
by means of a hydrodynamic approach based on the local LL model. 
The expansion dynamics was shown to be different for different interaction strengths, and its analysis 
could be employed to discern between the TG and MF regimes. In particular, the 
self-similar solution is violated. 

In this paper, we extend the analysis of Ref.\cite{Letter} by introducing a variational approach, which permits us 
to study the asymptotic regime at large expansion times. This method is shown to be in excellent agreement 
with previous  
direct numerical simulations, and additionally permits us to recover the results of Refs.\ \cite{Olshanii2} and 
\cite{Chiara}. More importantly, our variational approach allows us to determine the regime of parameters for which the 
self-similarity of the expansion is violated.

The paper is organized as follows. In Sec.\ \ref{sec:LLL} we introduce the local LL model which we employ to 
analyze the expansion dynamics. In Sec.\ \ref{sec:num} we briefly discuss the numerical results obtained in Ref.~\cite{Letter}. 
In Sec.\ \ref{sec:var} we present a variational approach which 
allows us to investigate in detail the expansion dynamics for arbitrary regimes of parameters. Finally we conclude in Sec.\ 
\ref{sec:concl}.

\section{Local Lieb-Liniger model}
\label{sec:LLL}

We analyze in the following a dilute gas of $N$ bosons 
confined in a very elongated harmonic trap with radial and axial frequencies 
$\omega_\rho$ and $\omega_z$ ($\omega_\rho\gg\omega_z$). 
We assume that the transversal confinement is strong enough so that the 
interaction energy per particle is smaller than the 
zero-point energy $\hbar\omega_\rho$ of the transversal trap. In this way, the 
transversal dynamics is effectively ``frozen'' and the 
system can be considered as dynamically 1D. 
In this section we briefly review the formalism introduced in Ref.\ 
\cite{Olshanii2}.

We assume that the interparticle interaction can be approximated by a delta function pseudopotential. Therefore 
the Hamiltonian that describes the physics of the 1D gas becomes
\begin{equation}
\hat{H}_{\rm 1D}=\hat{H}_{\rm 1D}^{0} + 
\sum_{j=i}^{N}\frac{m\omega_{z}^{2} z_{i}^{2}}{2}
\end{equation}
where
\begin{equation}
\hat{H}_{\rm 1D}^{0}=
-\frac{\hbar^{2}}{2m}\sum_{j=1}^{N}\frac{\partial^{2}}{\partial z_{j}^{2}}
+ g_{\rm 1D}\sum_{i=1}^{N-1}\sum_{j=i+1}^{N}\delta \left(z_{i}-z_{j}\right)
\end{equation}
is the homogeneous Hamiltonian in absence of the harmonic trap,  
$m$ is the atomic mass and $g_{\rm 1D}=-2\hbar^{2}/ma_{\rm 1D}$. 
The scattering problem under one-dimensional constraints was analyzed in detail 
by Olshanii \cite{Olshanii1}, and it is characterized by 
the one-dimensional scattering length 
$a_{\rm 1D}=(-a^{2}_{\rho}/2 a)[1-{\mathcal C}(a/a_{\rho})]$, 
with $a$ the three-dimensional scattering length, $a_\rho=\sqrt{2\hbar/m\omega_\rho}$   
the oscillator length in the radial direction, and ${\mathcal C}=1.4603\dots$.
As shown by Lieb and Liniger \cite{Lieb63}, 
the homogeneous Hamiltonian $\hat{H}_{\rm 1D}^{0}$ can be diagonalized exactly by means 
of Bethe Ansatz \cite{Kor}. In the thermodynamic limit, 
a 1D gas at zero temperature with a given linear density $n$, 
is characterized by the energy per particle
\begin{equation}
\epsilon(n)= \frac{\hbar^{2}}{2m}n^{2} e(\gamma(n)), 
\label{energie}
\end{equation}
where $\gamma = 2/n|a_{\rm 1D}|$. The function $e(\gamma)$ fulfills
\begin{equation}
e(\gamma) = \frac{\gamma^{3}}{\lambda^{3}(\gamma)}
\int_{-1}^{1}g\left(x|\gamma\right)x^{2}dx ,
\end{equation}
where $g\left(x|\gamma\right)$ and $\lambda(\gamma)$ are the solutions of
the LL system of equations \cite{Lieb63}
\begin{eqnarray}
g\left(x|\gamma\right)&=&\frac{1}{2\pi}+\frac{1}{2\pi}\int_{-1}^{1}
\frac{2\lambda(\gamma)}{\lambda^{2}(\gamma)+(y-x)^{2}}g\left(y|\gamma\right)dy
\label{g} \\
\lambda(\gamma)&=&\gamma \int_{-1}^{1}g\left(x|\gamma\right)dx.
\label{lambda}
\end{eqnarray}
We assume next that at each point $z$ the gas is in local 
equilibrium, with local energy per particle provided by Eq.\ (\ref{energie}). Then, one can 
obtain the corresponding hydrodynamic equations for the density and the 
atomic velocity
\begin{mathletters}
\begin{eqnarray}
\frac{\partial }{\partial t}n+\frac{\partial}{\partial z} (nv)&=&0 \label{hydn}\\
\frac{\partial }{\partial t}v+v\frac{\partial}{\partial z}v&=& -\frac{1}{m}
\frac{\partial}{\partial z}(\mu_{le}(n)+\frac{1}{2}m\omega_z^2 z^2).\label{hydv}
\end{eqnarray}
\end{mathletters}
where
\begin{equation}
\mu_{le}(n) = \left (1 + n\frac{\partial}{\partial n}\right )\epsilon(n)
\label{mule}
\end{equation}
is the Gibbs free energy per particle. 

Note that for the case of $n|a_{1D}|\rightarrow \infty$, one obtains 
$\mu_{le}(n)=g_{1D}n$, retrieving the 1D Gross-Pitaevskii equation (GPE) \cite{Gross}, whereas 
for the case $n|a_{1D}|\rightarrow 0$,  one gets $\mu_{le}(n)=\pi^2\hbar^2n^2/2m$, 
and the equation of Ref.\ \cite{Kolomeisky} is recovered. 
The system has only one control parameter \cite{Petrov1D,Olshanii2,Chiara}, 
namely $A=N |a_{1D}|^2/a_z^2$, where $a_z=\sqrt{\hbar/m\omega_z}$ is the harmonic oscillator length in the $z$ direction.
The regime $A\gg 1$ corresponds to the 
MF limit, in which the stationary-state density profile has a 
parabolic form. On the other hand, the regime $A\ll 1$ corresponds to the 
TG regime, which is characterized by a stationary-state density 
profile with the form of a square root of a parabola.

\section{Numerical results}
\label{sec:num}

\begin{figure}[ht] 
\begin{center}\ 
\psfig{file=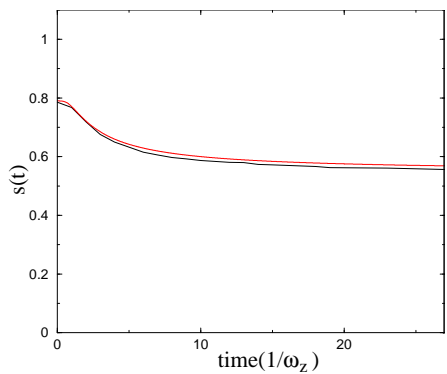,width=5.8cm}\\[0.1cm]
\end{center} 
\caption{
Time evolution of the exponent $s(t)$ for $A=0.43$, $\omega_\rho=2\pi\times 20$kHz and 
$N=200$ atoms ($\omega_z=2\pi\times 1.8$Hz at $t=0$). Our variational result (dashed line) 
shows a very good agreement with the results obtained from the direct resolution of 
Eqs.~(\ref{hydn}) and (\ref{hydv}).}
\label{fig:2}  
\end{figure}

In Ref.~\cite{Letter}, equations \ (\ref{energie}),(\ref{g}),(\ref{lambda}),(\ref{mule}) 
were employed to simulate numerically the expansion of a 1D gas 
in the framework of the hydrodynamic formalism. The expansion follows 
the sudden removal of the axial confinement, while 
the radial one is kept fixed. In particular, it was observed that during the expansion 
the density profile is well described by the expression 
\begin{equation}
n(z,t)=n_m(t)\left (1-\left (\frac{z}{R(t)}\right )^2\right )^{s(t)}
\label{app}
\end{equation}
where $n_m(t)$ provides the appropriate normalization, 
$R(t)$ is the radius of the cloud, and the exponent $s(t)$ takes the value $s(0)=1$ 
for an initial MF gas. The function $s(t)$ decreases monotonically in time, approaching an 
asymptotic value (see Fig.~\ref{fig:2}). Therefore, contrary to the expansion dynamics for a 
BEC \cite{Castin,Kagan1,Minniti}, the self-similarity of the density profile is violated. 

At this point we discuss the physics behind this violation of the self-similarity. If the 
local chemical potential presents a fixed power law dependence on the density, $\mu_{le}\propto n^\lambda$, 
it is easy to show from the hydrodynamic equations (\ref{hydn}) and (\ref{hydv}), 
that there exists a self-similar solution of the form 
$n=(n_0/b)(1-(z/bR)^2)^{1/\lambda}$, where $\ddot b=\omega_z^2/b^{\lambda+1}$. 
For the particular case of the TG gas, the scaling law can be also obtained from the exact 
BF mapping \cite{Letter}. However, since $\mu_{le}$ is obtained from 
the LL equations, the dependence of $\mu_{le}$ on $n$ is quadratic for a low density and linear for a large one. Therefore, 
$\mu_{le}$ does not fulfill a fixed power law dependence during the expansion, and the self-similarity 
is violated. In particular, as the expansion proceeds the whole system approaches the low density regime, 
and consequently the exponent $s$ decreases monotonically. 
In the next section, we analyze in more detail this effect.

\section{Variational calculation}
\label{sec:var}

In this section, we complete our understanding of the expansion of a one-dimensional 
Bose gas in a guide by means of a variational Ansatz using a Lagrangian formalism. 
The Lagrangian density for the system is of the form
\begin{eqnarray}
\mathcal{L}=-mn\frac{\partial \phi}{\partial t}-\frac{1}{2}mn
\left(
\frac{\partial \phi}{\partial z}
\right)^2-\frac{1}{2}m\omega_z^2z^2n-\varepsilon(n)n
\end{eqnarray}
where the velocity field is defined as $v=\partial \phi/\partial z$. 
The equations of motion are obtained from  
the functional derivation of the action ${\cal A}=\int \mathcal{L}dtdz$:  
$\delta {\cal A}/\delta\phi=0$ (continuity equation), 
$\delta {\cal A}/\delta n=0$ (which after partial derivation with respect to $z$ 
provides the Euler equation). From the numerical results we have observed that the 
density is at any time well described by Eq. (\ref{app}). Therefore, we assume the following 
Ansatz for the density 
\begin{equation}
n=\frac{C(s)}{b}
\left(1-\frac{z^2}{R^2b^2}
\right)^{s},
\label{den}
\end{equation}
where $b$ and $s$ are time dependent variables, $R$ is the 
initial Thomas-Fermi radius, and $C(s)$ is related to the normalization to
the total number of particles. For the $\phi$ field we consider the following form:
\begin{equation}
\phi=\frac{1}{2}\alpha z^2+\frac{1}{4}\beta z^4,
\label{phi}
\end{equation}
where $\alpha$ and $\beta$ are time dependent parameters. We stress at this point, that in the 
analysis of the self-similar expansion of a BEC \cite{Castin,Kagan1,Minniti}, 
a quadratic ansatz (in $z$) for the $\phi$ field provides the exact solution. However, 
for the problem under consideration in this paper, it is necessary to include higher order terms 
to account for the violation of the self-similarity. We have checked that 
terms of higher order than $z^4$ introduces only small corrections, and therefore we reduce to the 
form of Eq.~(\ref{phi}).

\begin{figure}[ht] 
\begin{center}\ 
\psfig{file=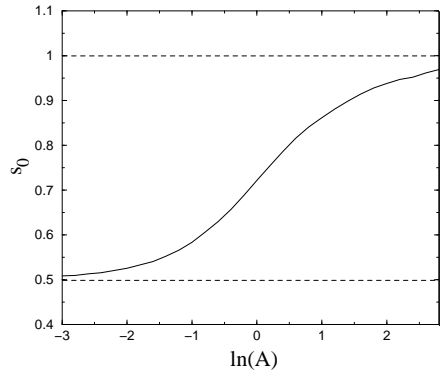,width=5.8cm,angle=0}\\[0.1cm]
\end{center} 
\caption{Equilibrium values (at $t=0$ before opening the trap) of the exponent $s$, 
as a function of the parameter $A$. The dashed 
lines denote the MF limit, $s_0=1$, and the TG one, $s_0=1/2$.}
\label{fig:gr}  
\end{figure}

\begin{figure}[ht] 
\begin{center}\ 
\psfig{file=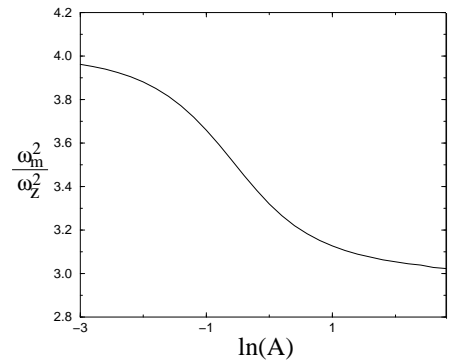,width=5.8cm}\\[0.1cm]
\end{center} 
\caption{Frequency of the lowest compressional mode as a function of the parameter $A$.}
\label{fig_wA}  
\end{figure}

We are interested in the dynamics of the parameters $b$ and $s$, related to 
the size and the shape of the cloud, respectively.
Integrating the Lagrangian density in $z$,  
$L=\int\mathcal{L}dz$, one finds a Lagrangian for the above mentioned parameters: 
\begin{eqnarray}
L(\dot{\alpha},&&\!\!\!\!\!\!\alpha,\dot{\beta},\beta,b,s) =\nonumber \\
&&\frac{mNR^2}{2}\left \{ -\frac{\dot{\alpha}b^2}{2s+3}-
\frac{3}{2}\frac{\dot{\beta}b^4}{(2s+5)(2s+3)} \right \delimiter 0 \nonumber\\
&&-\frac{\alpha^2b^2}{2s+3}-
2\frac{\alpha\beta b^4}{(2s+5)(2s+3)}-\nonumber \\
&&-\frac{\beta^2b^6}{(2s+7)(2s+5)(2s+3)} \nonumber \\
&&\left\delimiter 0 -\frac{b^2\omega_z^2}{2s+3} \right \}-\int dz n \epsilon (n).
\label{L1}
\end{eqnarray}
We perform a gauge transformation \cite{footnote2}
\begin{eqnarray}
L(t,q,\dot{q})\rightarrow L(t,q,\dot{q})+\frac{d}{dt}g(t,q),
\label{gauge}
\end{eqnarray}
where 
\begin{equation}
g(t,q)=\frac{mNR^2}{2}\left \{ -\frac{\alpha b^2}{2s+3}-
\frac{3}{2}\frac{\beta b^4}{(2s+5)(2s+3)} \right\}.
\end{equation}
The resulting Lagrangian is of the form  
$L=L(\alpha,\beta,\dot{b},b,\dot{s},s)$. Imposing the conservation laws 
$\partial L/\partial\alpha=\partial L/\partial\beta =0$, we obtain a Lagrangian 
depending only on the two relevant parameters $s$ and $b$, of the form $L=K-V$, where
\begin{eqnarray}
\frac{V}{E_{1D}}&=&\left(
\frac{B(b,s)}{f_0(s)}+\frac{A^2b^2}{[\eta_0f_0(s_0)]^2}
\frac{1}{2s+3}
\right), \label{V}\\
\frac{K}{E_{1D}}&=&\frac{A^2\left (
M_{11}\dot b^2+2M_{12}\dot b\dot s+M_{22}\dot s^2
\right )}{[\eta_0f_0(s_0)]^2 (2s+3)}
\label{K},
\end{eqnarray}
where $E_{1D}=\hbar^2/2m|a_{1D}|^2$ is the typical energy associated with the interatomic interactions. 
In Eq.~(\ref{V}), we use the function $B(b,s)=\int dy (1-y^2)^s \epsilon (n(y))/E_{1D}$, where we integrate over the 
rescaled axial coordinate $y=z/Rb$. In Eqs.~(\ref{V}) and (\ref{K}), we define the dimensionless central density
$\eta_0=n_0|a_{1D}|$, where $n_0$ is the initial central density, and the parameter   
$s_0=s(t=0)$. We have additionally employed the auxiliary functions 
$f_n(s)=\int y^n (1-y^2)^s dy$, and the coefficients
\begin{eqnarray}
M_{11}&=&1, \\
M_{12}&=&\frac{-b}{2s+3} \\
M_{22}&=&\frac{b^2(121+186s+96s^2+16s^3)}{4(s+1)(2s+3)^2(2s+5)^2}
\end{eqnarray}

From the Lagrangian $L$, one obtains the corresponding Euler-Lagrange equations for the parameters $b$ and $s$.  
In order to find the initial conditions $\eta_0$ and $s_0$ for the expansion, 
we have numerically minimized the potential $V$ in 
the presence of the harmonic trap for different values of $A$, assuming $b=1$ (see Fig.\ \ref{fig:gr}). 
When $A\gg 1$, $s_0$ tends to $1$, as expected for the MF case. On the contrary,  
when $A\ll 1$, $s_0$ tends to $1/2$ (TG profile).
As expected from Ref.\ \cite{Olshanii2}, 
for $A\ll 1$, $\eta_0\propto A^{1/2}$, whereas for $A\gg 1$, the MF dependence $\eta_0\propto A^{2/3}$ is recovered. 

Our variational approach allows us to calculate the lowest compressional mode, offering an alternative method as
the one discussed in Ref.~\cite{Chiara}. Expanding the potential $V$ around the 
equilibrium solution up to second order in $b$ (see Fig.\ \ref{fig_wA}), and neglecting   
for small oscillations the time-dependence of $s$, we obtain
\begin{eqnarray}
\frac{\omega_m^2}{\omega_z^2}=1+\frac{1}{2A^2}
\frac{[\eta_0^2f_0(s_0)]^2}{f_2(s_0)}
\left\delimiter 0 \frac{\partial^2 B}{\partial b^2} \right |_{b=1}.
\end{eqnarray}
Our results show a continuous transition from the MF 
value, $\omega_m=\sqrt{3}\omega_z$, to the TG one, $\omega_m=2\omega_z$, in excellent agreement
with the results obtained by means of a sum rule formalism \cite{Chiara}.

\begin{figure}[ht] 
\begin{center}\ 
\psfig{file=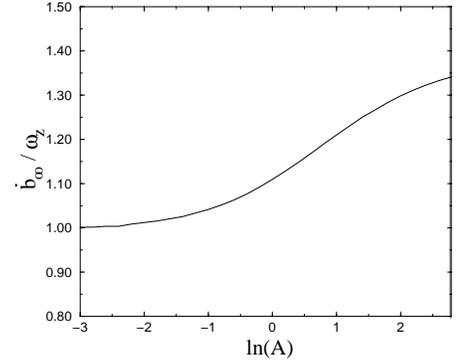,width=5.8cm}\\[0.1cm]
\end{center} 
\caption{Expansion velocity. 
Asymptotic value of $\dot b$ as a function of the parameter $A$.
}
\label{fig:binf}  
\end{figure}

\begin{figure}[ht] 
\begin{center}\ 
\psfig{file=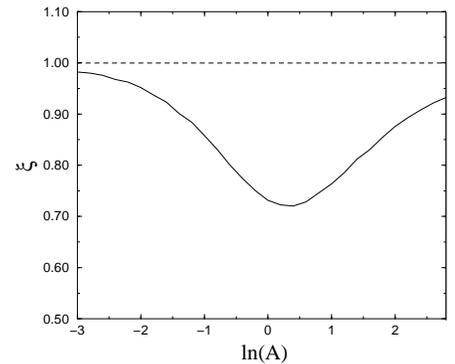,width=5.8cm}\\[0.1cm]
\end{center} 
\caption{Violation of the self-similarity.  
Value of $\xi=s_\infty/s_0$ as a function of the parameter $A$. The value $\xi=1$ denotes self-similarity.}
\label{fig:sinf}  
\end{figure}

From the corresponding Euler-Lagrange equations, we have obtained the dynamics of $b(t)$ and $s(t)$. We 
have checked in all our calculations that the energy and number of particles remain a constant of motion.  
We have compared the variational results with our simulations based on the exact resolution of the 
hydrodynamic equations \cite{Letter}, obtaining an excellent agreement (see Fig.\ \ref{fig:2}). 

We have analyzed the asymptotic value, $\dot b_\infty$, for different values of $A$. It is easy to obtain, that 
for a power law dependence of the local chemical potential $\mu_{le}\propto n^\lambda$, 
the derivative of the scaling parameter asymptotically approaches a value  
$\dot b_\infty=\sqrt{2/\lambda}\omega_z$. Therefore a continuous transition from $\dot b_\infty=\omega_z$ (TG) 
to $\dot b_\infty=\sqrt{2}\omega_z$ (MF) is expected. We recover this dependence from our variational 
calculations (see Fig.\ \ref{fig:binf}). 

We have analyzed the behavior of $s$ during the expansion dynamics for different values of $A$. In particular, 
we have defined the asymptotic ratio $\xi=s_\infty/s_0$, with $s_\infty=s(t\rightarrow\infty)$ (see Fig.\ \ref{fig:sinf}). 
Deeply in the TG regime ($A\ll 1$) or in the MF one ($A\gg 1$), $\xi\simeq 1$, i.e. in those 
extreme regimes, the expansion is well-described by a self-similar solution. However, for intermediate values, 
$\xi<1$, i.e. the expansion is not self-similar. The self-similarity is maximally violated in the vicinity 
of $A=1$, although $\xi$ departs significantly from $1$ for a range $0.01\lesssim A \lesssim 100$.
The behavior at large $A$ can be understood as follows. If the gas is at $t=0$ 
deeply in the MF regime, the change of the functional dependence of $\mu_{le}$ with the density 
occurs at very long expansion times, when the initial interaction energy of the gas has been 
fully transferred into kinetic energy. Therefore, for large values of $A$ the self-similarity is recovered.

\section{Conclusions} 
\label{sec:concl}

In this paper, we have extended the analysis of Ref.~\cite{Letter} 
on the expansion dynamics of a one-dimensional Bose gas in a guide. 
We have shown that the expansion violates under certain conditions the self-similarity, and in this sense differs significantly 
from the expansion dynamics of a BEC. We have shown that the problem can be solved 
by employing the hydrodynamic approach, and the local Lieb-Liniger model.
We have developed a variational approach based on a Lagrangian formalism to study the expansion for any regime of 
parameters. We have identified the possible physical situations at which self-similarity is violated. This should occur in 
a rather wide range of parameters ($0.01\lesssim A\lesssim 100$). The particular properties of the expansion of a gas 
in the strongly-interacting regime could therefore be employed to discern between mean-field and strongly-interacting regimes. 
In addition, the asymptotic behavior of the expanded cloud could be employed to discriminate between different initial 
interaction regimes of the system.
 
Our discussion has been restricted to the analysis of the density properties. In fact the present formalism cannot describe the 
dynamics of the coherence in the system, i.e. we are limited to the diagonal terms of the corresponding 
single-particle density matrix. The description of the non diagonal terms lies beyond the scope of this paper, and 
requires other techniques of analysis \cite{Gangardt,Giorgini}. 


We acknowledge  support from Deutsche Forschungsgemeinschaft (SFB 407),  
the RTN Cold Quantum gases, ESF PESC BEC2000+, Royal Society of Edinburgh, and the Ministero dell'Istruzione, 
dell'Universit\`a e della Ricerca (MIUR).
L. S. and P. P. wish to thank the Alexander von Humboldt Foundation, 
the Federal Ministry of Education and 
Research and the ZIP Programme of the German Government.
Discussions with M. D. Girardeau, 
M. Lewenstein, D. S. Petrov, and G. V. Shlyapnikov are acknowledged. 
We especially thank C. Menotti for providing us with the data of Ref.\ \cite{Chiara}.


\begin{references}
\bibitem{footnote2} The Lagrangian (\ref{L1}) does not depend on $\dot{b}$ and $\dot{s}$, and consequently one obtains the 
dynamics only for $\alpha$ and $\beta$ imposing $\partial L/\partial b
=\partial L/\partial s=0$, and later on from these two conditions one can obtain
the dynamics of $b$ and $s$. The gauge transformation (\ref{gauge}) turns out to 
simplify significantly the calculations.
\end{references}
\end{document}